\documentclass[sigconf]{acmart}

\usepackage{amsmath}

\AtBeginDocument{%
  \providecommand\BibTeX{{%
    \normalfont B\kern-0.5em{\scshape i\kern-0.25em b}\kern-0.8em\TeX}}}

\usepackage{bm}

\usepackage{amsfonts}
\usepackage{amsmath}
\usepackage{booktabs}
\usepackage{algorithmic}
\usepackage{graphicx}
\usepackage{subfigure}
\usepackage[boxed,ruled,lined]{algorithm2e}
\usepackage{amsthm}
\usepackage[american]{babel}
\usepackage{setspace}
\usepackage{enumerate}
\usepackage{multirow}
\usepackage{url}
\usepackage{hyperref}
\hypersetup{colorlinks, linkcolor=blue}
\usepackage{arydshln} 
\usepackage{amssymb}

\newcommand{\argmin}{\operatornamewithlimits{arg\,min}}

\begin{document}

\copyrightyear{2022}
\acmYear{2022}
\setcopyright{acmcopyright}
\acmConference[WWW '22] {Proceedings of the ACM Web Conference 2022}{April 25--29, 2022}{Virtual Event, Lyon, France.}
\acmBooktitle{Proceedings of the ACM Web Conference 2022 (WWW '22), April 25--29, 2022, Virtual Event, Lyon, France}
\acmPrice{15.00}
\acmISBN{978-1-4503-9096-5/22/04}
\acmDOI{10.1145/3485447.3511951}

\title{A Model-Agnostic Causal Learning Framework for Recommendation using Search Data}

\author{
        Zihua Si\textsuperscript{\rm 1,\rm 2},
        Xueran Han\textsuperscript{\rm 1,\rm 2},
        Xiao Zhang\textsuperscript{\rm 1,\rm 2},
        Jun Xu\textsuperscript{\rm 1,\rm 2,*},
        Yue Yin\textsuperscript{\rm 3},
        Yang Song\textsuperscript{\rm 3},
        Ji-Rong Wen\textsuperscript{\rm 1,\rm 2}
}
\affiliation{
    \textsuperscript{\rm 1} Gaoling School of Artificial Intelligence, Renmin University of China
    \city{Beijing} \country{China}; \\
    \textsuperscript{\rm 2} Beijing Key Laboratory of Big Data Management and Analysis Methods;
    \textsuperscript{\rm 3} Kuaishou Technology Co., Ltd., China \\
    \{2018202181, hanxueran, zhangx89, junxu, jrwen\}@ruc.edu.cn; \{yinyue, yangsong\}@kuaishou.com\\
}

\thanks{* Corresponding author: Jun Xu}

\begin{abstract}

Machine-learning based recommender system(RS) has become an effective means to help people automatically discover their interests. 
Existing models often represent the rich information for recommendation, such as items, users, and contexts, as embedding vectors and leverage them to predict users' feedback.
In the view of causal analysis, the associations between these embedding vectors and users' feedback are a mixture of the causal part that describes why an item is preferred by a user, and the non-causal part that merely reflects the statistical dependencies between users and items, for example, the exposure mechanism, public opinions, display position, etc.
However, existing RSs mostly ignored the striking differences between the causal parts and non-causal parts when using these embedding vectors.
In this paper, we propose a model-agnostic framework named IV4Rec that can effectively decompose the embedding vectors into these two parts, hence enhancing recommendation results.
Specifically, we jointly consider users' behaviors in search scenarios and recommendation scenarios. 
Adopting the concepts in causal analysis, we embed users' search behaviors as \emph{instrumental variables} (IVs), to help decompose original embedding vectors in recommendation, i.e., \emph{treatments}.
IV4Rec then combines the two parts through deep neural networks and uses the combined results for recommendation.
IV4Rec is model-agnostic and can be applied to a number of existing RSs such as DIN and NRHUB. 
Experimental results on both public and proprietary industrial datasets 
demonstrate that IV4Rec consistently enhances RSs and outperforms a framework that jointly considers search and recommendation.

\end{abstract}

\ccsdesc[500]{Information systems~Recommender systems}

\keywords{recommendation; search; causal learning; instrumental variables}

\maketitle

{\fontsize{8pt}{8pt} \selectfont
\textbf{ACM Reference Format:}\\
Zihua Si, Xueran Han, Xiao Zhang, Jun Xu, Yue Yin, Yang Song, Ji-Rong Wen. 2022. A Model-Agnostic Causal Learning Framework for Recommendation using Search Data. In {\it Proceedings of the ACM Web Conference 2022 (WWW ’22), April 25–29, 2022, Virtual Event, Lyon, France.} ACM, New York, NY, USA, 9 pages. \url{https://doi.org/10.1145/3485447.3511951}}

\section{Introduction}
Recommendation and search have become two major approaches to help users to obtain information from the Internet. Traditionally, recommendation and search were usually deployed as two separate systems, serving different users with different types of information objectives. In recent years, many online content platforms provide both search and recommendation services in one application. While having heterogeneous user inputs, these two services can be connected through their common sets of users and items. This phenomenon provides us the opportunity to further improve the performance of one service through using the user activities collected from the other. Early studies have been conducted and showed that jointly optimizing the search and recommendation can improve their respective performances ~\cite{Zamani2020Learning,zamani2018joint}.

Traditional recommender systems(RSs) utilize the rich information from the user, the item, and the context to make recommendations. Usually, this information is represented as real-valued embedding vectors. The users' preference to an item, therefore, is calculated based on these embeddings, e.g., using dot product between the user and item embeddings. From the viewpoint of causal analysis, the signals characterized by the embeddings can be decomposed into two parts: the causal association part which describes why a user prefers an item under the context; the non-causal association part, on the other hand, is often affected by many factors, such as the exposure mechanism, public opinions, display position, etc. Thus it merely reflects the statistical dependencies between users and items. The striking differences between causal and non-causal associations lead to their different roles in RSs. While the causal association part contains key signals that lead to the outcomes (e.g., clicks), the non-causal association part may still influence the outcomes through a few unobserved confounders. 

Since the causal association part and non-causal association part in embedding vectors affect the final recommendation through different mechanisms, to achieve optimal performance, an ideal RS can employ different approaches to handle corresponding signals respectively. However, existing RSs mostly ignored the differences between the two parts through using the embeddings as a whole. In this paper, we propose a model-agnostic framework named IV4Rec that can effectively decompose the embedding vectors into these two parts by jointly considering users' behavior in search scenarios and recommendation scenarios. Specifically, adopting the concepts in causal analysis, we embed users' search behavior as \emph{instrumental variables} (IVs), to help decompose original embedding vectors in recommendation, i.e., \emph{treatments}.

In causal inference, IVs methods have been widely used to predict the effects of unobserved causes~\cite{Jason2017Deep}.
After identifying the IVs that only affect the treatments and not the confounders, the IV regression can basically split the treatment variable into two parts: one part that has causal correlation and one part that probably does not. 
Since the search and recommendation services are deployed in one platform and are with shared user groups and candidate items, users' search activities also reflect their preferences in recommendation scenarios. Therefore, it is reasonable to take users' search activities as IVs to decompose the recommendation embeddings into causal association and non-causal association parts. 

In our framework, when considering to recommend an item to a user, a set of queries related to this item is collected as IVs for this item. For example, queries that most users search for before clicking on this item. The IVs are represented by embeddings and are used to fit the original embedding of this item through a regression model. In this way, the original item embedding can be successfully decomposed into the causal part (the values fitted by the regression model) and the non-causal part (the residuals). Finally, these two parts are reconstructed into a new vector (new treatment) and fed into the RS. Since users are usually represented by their browsing histories in RSs, the embedding of each item in a user's history can also be decomposed in this way, hence further enhancing the representation of users. 

The contributions of this paper are summarized as follows:

(1) We propose a model-agnostic framework, IV4Rec, to improve recommendation using search data. By considering users' search behavior as IVs to help decompose original embeddings in RS, the framework is able to enhance the representation of both users and items in RSs.

(2) We propose an approach to constructing new treatments through original embeddings and IVs. We use a regression model to decompose original embeddings into a causal part and a non-causal part, which are combined using neural networks and can be jointly trained with any suitable RSs.

(3) We conducted extensive experiments on a public dataset and a real-world industrial dataset. Experimental results demonstrated that IV4Rec can consistently enhance different RSs. In particular, using search activities as IVs for recommendation outperformed traditional methods that jointly model search and recommendation but ignore the causal effect.

\section{Related Work}
Traditionally, search and recommendation are designed as two separate systems and a large number of search models~\cite{Croft2009Search} and RSs~\cite{Ricci2011Recommender} have been developed. 
\citet{garcia2011information} also pointed out that search (information retrieval) and recommendation (information filtering) are the two sides of the same coin. They have strong connections and similarities~\cite{xu2018deep}. Recently, there is a trend to jointly model and optimize the search and recommendation and promote their accuracy at the same time~\cite{Zamani2020Learning,zamani2018joint}. For example, \citet{zamani2018joint} assume that search engines and RSs could potentially benefit from each other and designed a joint learning framework;~\citet{Zamani2020Learning} extend the work by joint learning search and recommendation models from user-item interactions; \citet{Jing2021USER} design an approach called USER that mines user interests from the integrated sequence and accomplishes these two tasks in a unified way, and applied to the tasks of personalized search and recommendation.

Besides the joint modeling, methods also have been developed to make use of search or recommendation as the external information to improve the performances of recommendation and search~\cite{Wu2020Zero-Shot,yao2012product,wu-etal-2019-neural}.~\citet{Wu2020Zero-Shot} propose a Zero-Shot Heterogeneous Transfer Learning framework that transfers the learned knowledge from the recommendation component to the search component, addressed the cold-start problem in the search system. \citet{yao2012product,wu-etal-2019-neural} use the search history log to enhance the recommendation task as external information. In this paper, we also make use of the search data as external information to enhance the recommendation.

In this paper, we use the instrumental variables (IVs)~\cite{caner2004instrumental,Venkatraman2016Online,Hartford2017Deep}, a popular method in causal inference~\cite{Imbens2015Causal}, to enhance recommendation with search data. 
Most IVs works make use of a two-stage least squares (2SLS) procedure~\cite{kmenta2010mostly}. 
Recently many IV-based causal learning methods extend 2SLS with deep learning methods. 
\citet{Jason2017Deep} provide a flexible framework to combine deep learning methods and the 2SLS method. 
\citet{xu2021learning} provide an alternating training regime for 2SLS and attain good end-to-end performance in high dimensional image data and off-policy reinforcement learning tasks. 
\citet{yuan2022autoIV} utilize mutual information to learn IV representation and confounder representation, which are used as inputs for two-stage regression with neural networks structure. 
In recommendation, causal learning has been used for tackling problem of the biases (e.g., position bias, popularity bias, selection bias etc.)~\cite{su2009survey,Agarwal2019Estimating,Ovaisi2020Correcting,Narita2021Debiased} and fairness~\cite{Kusner2017Counterfactual,Geyik2019Fairness,Morik2020Controlling}. Many researchers focus on causal embedding for recommendation \cite{bonner2018causal, zheng2021disentangling, wu2021learning, islam2021debiasing}. They are interested in finding the optimal treatment recommendation policy that maximizes the reward concerning the control recommendation policy for each user \cite{bonner2018causal} or learning a fairness or unbiased representation of items and users for recommendation\cite{zheng2021disentangling, wu2021learning,islam2021debiasing}. Other researchers propose a few methods to fit the preference of users with weighted click data, where each click is weighted by the inverse probability (IPW) of exposure \cite{liang2016causal, wang2020causal, schnabel2016recommendations,Zhang2021Counterfactual}.

\section{Problem Formulation}
This section formalizes the problem of recommendation with search queries as IVs. 
\subsection{Background}
\subsubsection{Recommendation and search in one platform}
A number of content platforms provide both search and recommendation, which serve the same set of users with the same set of items. From the viewpoint of recommendation, when a user $u\in \mathcal{U}$ accesses the platform, the system provides a list of items $i\in \mathcal{I}$ with an existing RS. 
Often, user $u$ interacts with items in certain context denoted as $\mathbf{p}_u$, including the user profile, search history, or situational context, which can be collected by the platform and represented as real-valued vectors (embeddings) $\mathbf{p}_u \in \mathbb{R}^{d_c}$, where $d_c$ is the dimension of embedding for context. 
Usually, each user $u$ and each item $i$ can also be represented as real-valued vectors (embeddings), denoted as $\mathbf{t}_u \in \mathbb{R}^{d_u}$ and $\mathbf{t}_i \in \mathbb{R}^{d_i}$, respectively, where $d_i$ and $d_u$ are the dimensions of the embeddings for users and items. 
The RS is usually trained with the historical user-system interactions $\mathcal{D}^{\mathrm{rec}}$ where each tuple $(u, i, c)\in \mathcal{D}^{\mathrm{rec}}$ means that the item $i$ was shown to the user $u$ and the interaction is $c\in\{0, 1\}$ where $c=1$ means clicked and $c=0$ otherwise. 

From the viewpoint of search, when a user $u\in\mathcal{U}$ issues a query $q\in \mathcal{Q}$ where $q$ is a text query and $\mathcal{Q}$ is the set of all queries, the system also provides a list of items $i\in\mathcal{I}$ with an existing search model. Similarly, each query can be represented as an embedding vector $\mathbf{t}_q \in \mathbb{R}^{d_q}$, where $d_q$ is the embedding dimension. Since search and recommendation shared the same set of users $\mathcal{U}$ and items $\mathcal{I}$, the user $u$ and item $i$ in search are also 
represented as the same embeddings $\mathbf{t}_u$ and $\mathbf{t}_i$ which are identical to those in recommendation. The historical user-system interactions in the search can be denoted $\mathcal{D}^{\mathrm{src}}$ where each tuple $(u, q, i, c) \in \mathcal{D}^{\mathrm{src}}$ indicates that a user $u$ is shown with item $i$ after issuing the query $q$, and the user's activity is $c \in \{0, 1\}$.
\begin{figure}
        \centering
            \includegraphics[width = 0.46\textwidth]{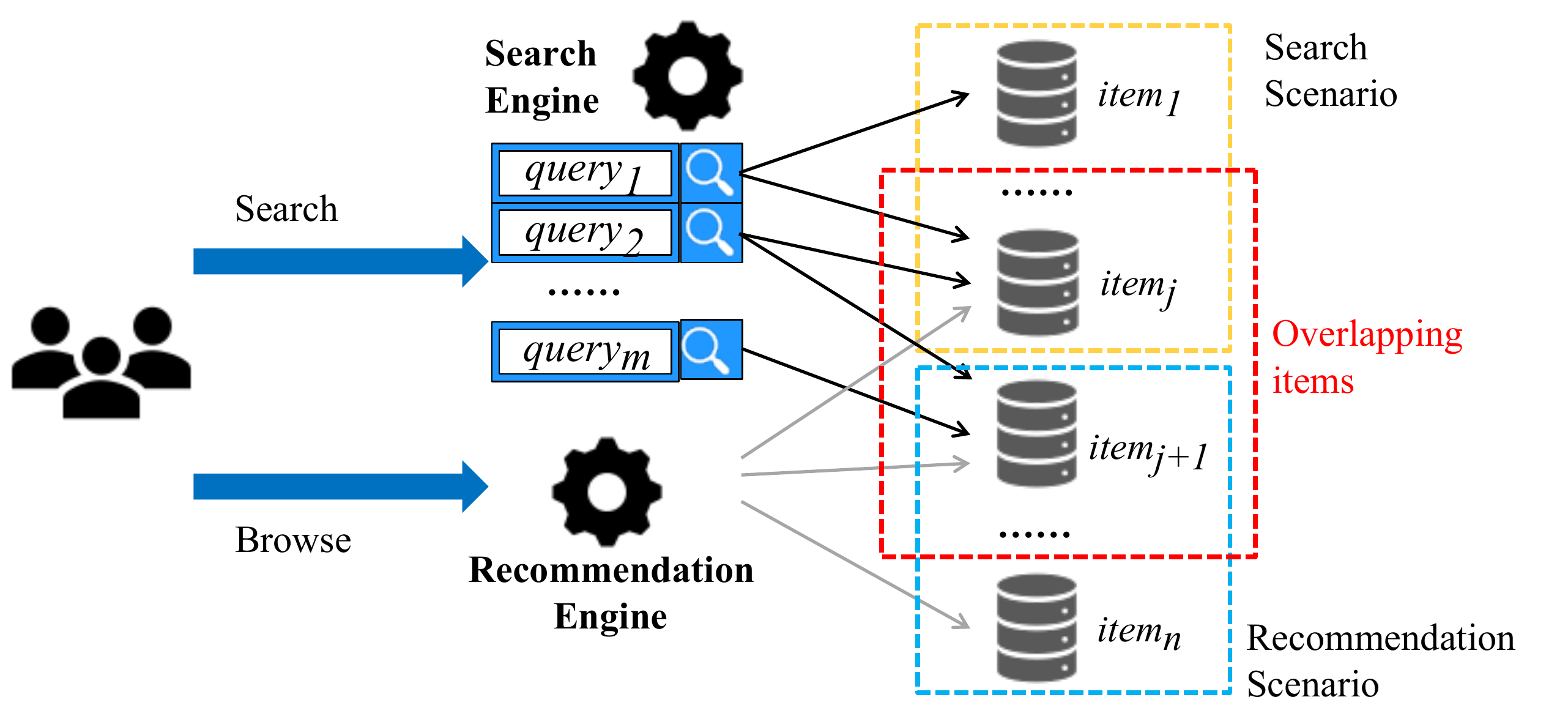}
        \caption{Recommendation and search services in one platform. Search scenario: users issue queries and click returned items. Recommendation scenario: users browse returned items. There exist overlapping items in both services.} 
\label{fig:overlap}
\vspace{-0.5cm}
\end{figure}
Since the search and recommendation serve the users with the same set of items, it is inevitable that there exist overlaps between $\mathcal{D}^{\mathrm{rec}}$ and $\mathcal{D}^{\mathrm{src}}$, that is, they have common target items in their records. As shown in~\autoref{fig:overlap}, there exist overlapping items in search and recommendation scenarios. 

\subsubsection{Method of instrumental variables} 
In causal inference, the method of IVs~\cite{Chernozhukov2007Instrumental,Belloni2012Sparse} aims to estimate the causal effect between a treatment variable $X$ and an outcome variable $Y$, in the presence of other variables (e.g., confounders) that are associated with the treatment and outcome simultaneously. Theoretically, a variable $Z$ is a valid \emph{instrumental variable} if it is unconfounded by the confounder (may be unobserved) and only affects the outcome $Y$ via the treatment $X$. Typical IVs methods such as 2SLS~\citep{kmenta2010mostly} adopt a two-stage least square regression to find the causal effect of treatment $X$ on the outcome $Y$: first regresses the treatment on the instrument and obtains a reconstructed treatment; then regresses the outcome on the reconstructed treatment from the first stage. An unbiased estimate of causal effect can be achieved from the coefficients of the second stage regression. 

\subsection{Causal view of recommendation}
\begin{figure}
\subfigure[Conventional recommendation models mix the causal and non-causal associations between treatment and outcome.]{\includegraphics[width=0.20\textwidth]{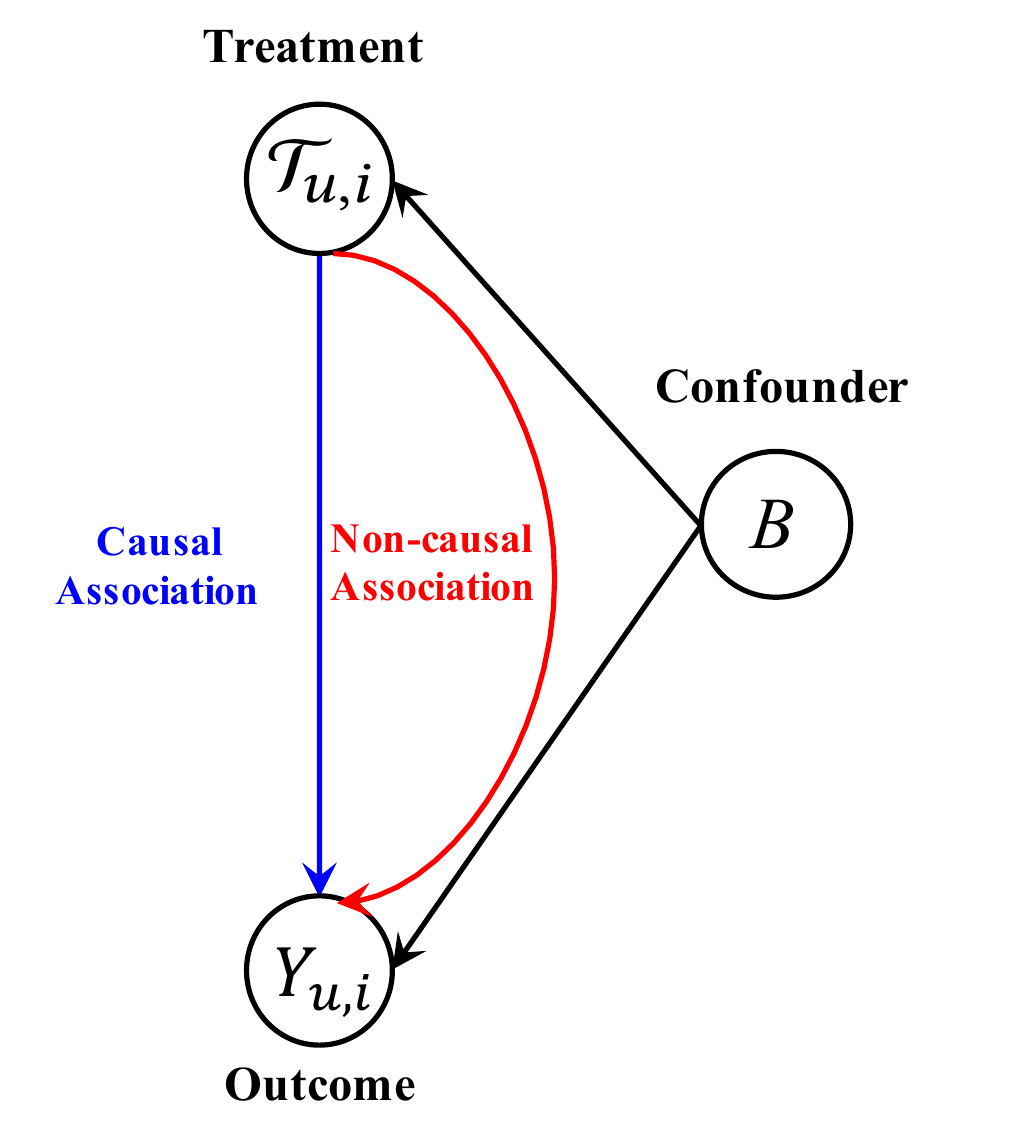}}
\hspace{1mm}
\subfigure[IV4Rec reconstructs treatment by leveraging IVs to decompose treatment into causal and non-causal parts and combining them with different weights.  ]{\includegraphics[width=0.22\textwidth]{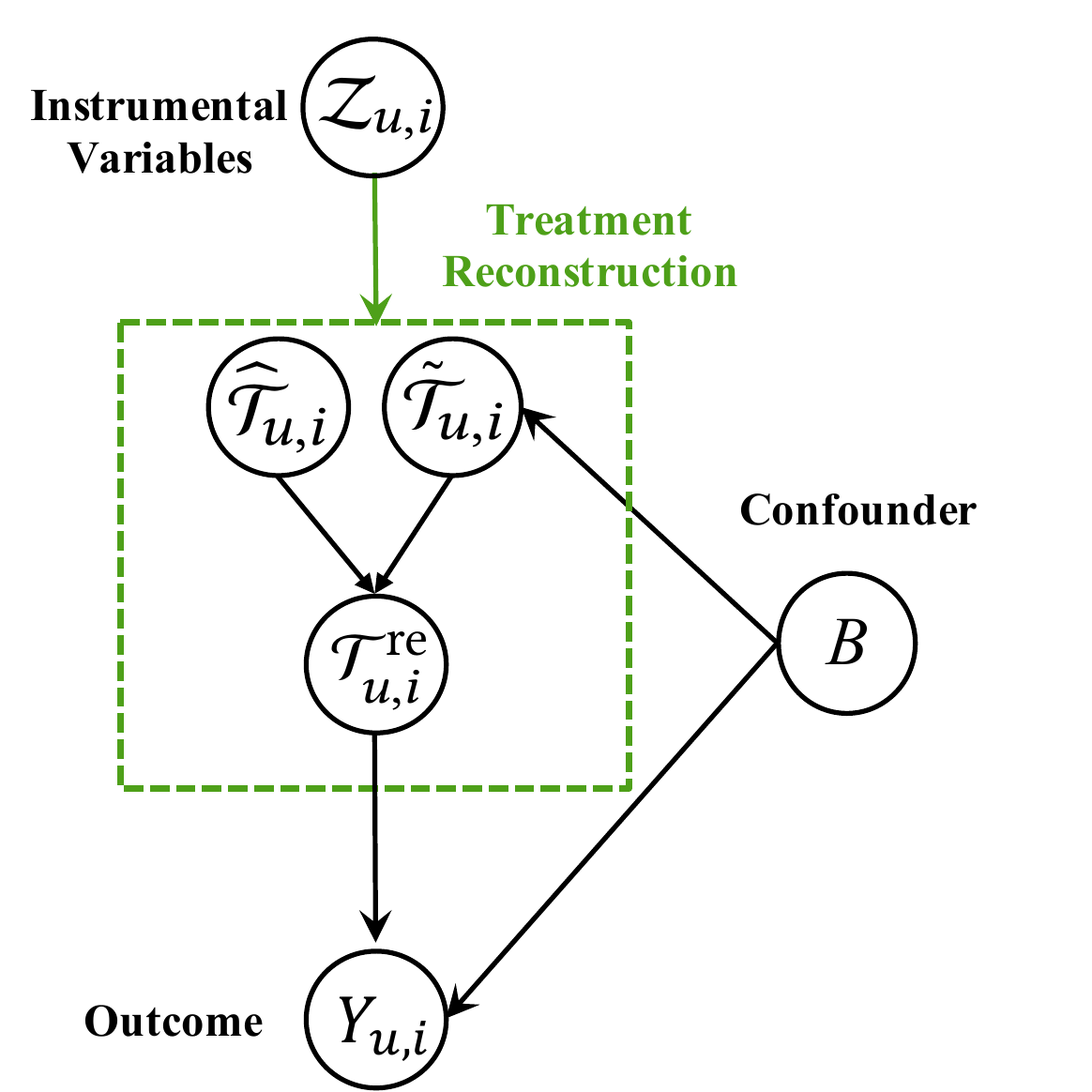}}
\caption{(a): conventional RSs. (b): RSs intervened by IVs. $\mathcal{T}_{u,i}$: embeddings of user and item. 
$B$: confounders (e.g. popularity bias, position bias). $Y_{u,i}$: user feedback. 
$\mathcal{Z}_{u,i}$: IVs (i.e. queries). 
$\widehat{\mathcal{T}}_{u,i}$: the fitted vectors. 
$\tilde{\mathcal{T}}_{u,i}$: the residuals.}
\label{fig:overall}
\vspace{-0.5cm}
\end{figure}
Existing RSs are usually trained on the user-system historical activities $\mathcal{D}^{\textrm{rec}}$, with the assumption that the click $c$ in each of the training records $(u, i, c) \in \mathcal{D}^{\textrm{rec}}$ unbiasedly reflects the preference of $u$ to $i$. In real world, however, the user clicks recorded in $\mathcal{D}^{\textrm{rec}}$ can be often affected by many factors (e.g., confounders), including the position bias, selection bias~\citep{popularity_bias@2017adressa}, and popularity bias~\citep{hernandez2014probabilistic}, etc. 
From the viewpoint of causal inference, 
we can regard the embedding vectors (i.e., the representations of users and items) as the treatment $\mathcal{T}_{u,i}$, and the user's feedback (i.e., click) as the outcome $Y_{u,i}$. 

Following the framework in~\cite{Pearl2009Causality}, a causal graph 
for conventional RSs can be constructed in \autoref{fig:overall}(a), where conventional RSs simply estimate the mixed associations between the treatment $\mathcal{T}_{u,i}$ and the outcome $Y_{u,i}$.
Due to the presence of (unknown) confounders $B$, there exist two paths from treatment $\mathcal{T}_{u,i}$ to outcome $Y_{u,i}$, including a path of non-causal association that is facilitated by the confounder (the red arrow curve from $\mathcal{T}_{u,i}$ to $Y_{u,i}$), and a path of causal association that describes why an item is preferred by a user (the blue arrow line from $\mathcal{T}_{u,i}$ to $Y_{u,i}$). 
Specifically, the non-causal association part is affected by confounders, such as the exposure mechanism, public opinions, display position, etc. Thus non-causal and causal associations reflect different relations between user-item pair (i.e., treatments) and user's feedback (i.e., outcome). 

It is difficult to identify the causal associations based on the biased observations $\mathcal{D}^{\textrm{rec}}$ given the unknown number of unknown confounders. Fortunately, the users' search activities in $\mathcal{D}^{\mathrm{src}}$ provide us a chance to decompose the treatment $\mathcal{T}_{u,i}$. 
As shown in \autoref{fig:overall}(b), we leverage the related queries as IVs, denoted as $\mathcal{Z}_{u,i}$, and regress $\mathcal{T}_{u,i}$ on $\mathcal{Z}_{u,i}$ to get $\widehat{\mathcal{T}}_{u,i}$ which doesn't depend on the confounders $B$. Thus the relation between $\widehat{\mathcal{T}}_{u,i}$ and $Y_{u,i}$ can be seen as a  causal association. We also calculate the residuals $\tilde{\mathcal{T}}_{u,i}$ of the regression. The relation between $\tilde{\mathcal{T}}_{u,i}$ and $Y_{u,i}$ can be seen as a non-causal association. 
Treatments are reconstructed by combining the fitted vectors $\widehat{\mathcal{T}}_{u,i}$ and the residuals $\tilde{\mathcal{T}}_{u,i}$. Therefore users' search activities are injected into RSs under a causal learning framework.

\section{Our approach: IV4Rec}
This section describes the proposed IV4Rec framework.

\subsection{Model overview}
\begin{figure*}
\includegraphics[width=0.90\textwidth]{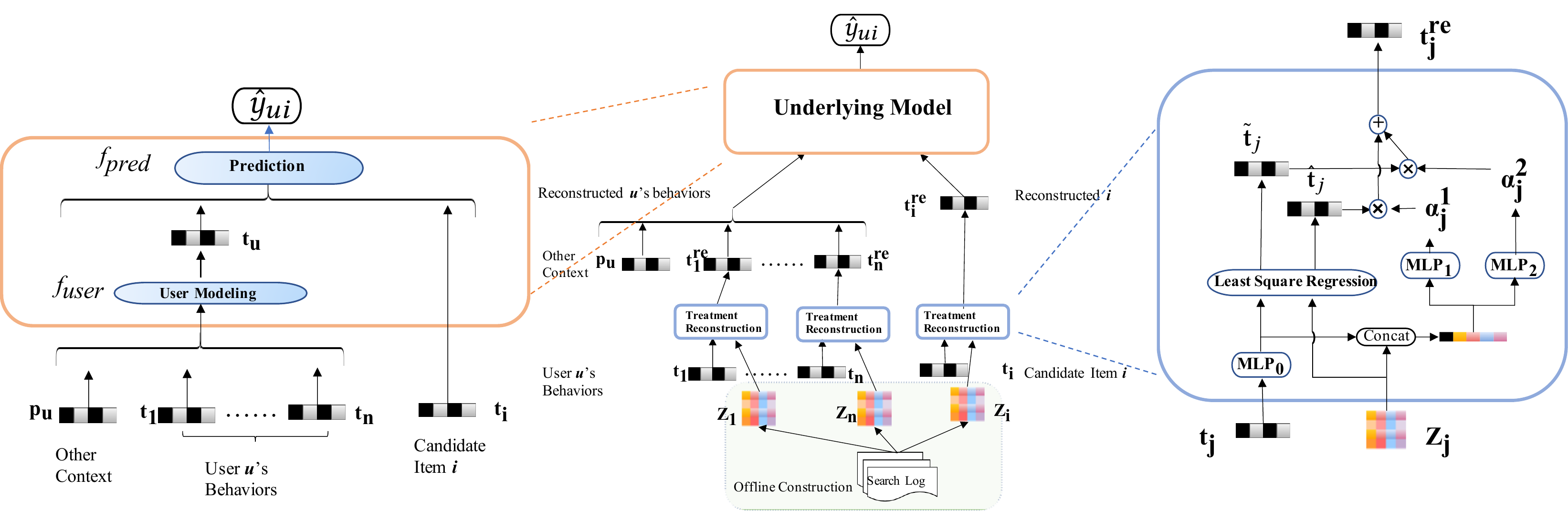}
\caption{The architecture of IV4Rec framework. 
Middle: the procedure of IV4Rec applied to the underlying model. Left: structure of the underlying model. 
Right: detailed implementation of treatment reconstruction. 
}
\label{fig:framework}
\end{figure*}

IV4Rec mainly consists of three steps, shown in~\autoref{fig:framework}. First, it defines the treatment $\mathcal{T}_{u,i}$ based on the recommendation data $\mathcal{D}^{\textrm{rec}}$, and constructs the IVs $\mathcal{Z}_{u,i}$ based on the search data $\mathcal{D}^{\mathrm{src}}$. Then, it reconstructs the treatment $\mathcal{T}^{\mathrm{re}}_{u,i}$ through regressing treatment $\mathcal{T}_{u,i}$ on IVs $\mathcal{Z}_{u,i}$. Finally, the reconstructed treatments are fed to a RS.

\subsection{Construction of treatments and IVs}
\label{const_V_Z}
To predict the preference score of a target user-item pair $(u, i)$, 
a treatment variable $\mathcal{T}_{u,i}$ in RS can be defined as a set of embeddings, including the embedding of the target item $i$ and the embeddings of the items interacted with $u$:
\begin{equation}\label{eq:treatment}
\mathcal{T}_{u,i} = \{\mathbf{t}_j:j\in \mathcal{I}_{u}\cup \{i\}\},
\end{equation}
where $\mathbf{t}_j \in \mathbb{R}^{d_i}$ is the embedding vector of item $j$, which is usually generated by some representation learning methods that project features (e.g., content) into a dense vector, and $\mathcal{I}_{u}$ denotes the set of items interacted with the user $u$ in $\mathcal{D}^{\textrm{rec}}$:
\[
\mathcal{I}_{u}=\left\{i':\exists (u, i', c=1)\in \mathcal{D}^{\mathrm{rec}}\right\}.
\]

The corresponding IVs $\mathcal{Z}_{u,i}$ of treatment $\mathcal{T}_{u,i}$ is defined as a set of matrices $\mathbf{Z}_j$: 
\begin{equation}\label{eq:IV}
\mathcal{Z}_{u,i} = \{\mathbf{Z}_j:j\in \mathcal{I}_{u}\cup \{i\}\},
\end{equation}
where each matrix $\mathbf{Z}_j$ is defined as a stack of the embeddings of the search queries related to item $j$. Please note that each $\mathbf{Z}_j$ corresponds to the vector $\mathbf{t}_j$ in treatment $\mathcal{T}_{u,i}$. 
Specifically, $\mathbf{Z}_j$ can be constructed as follows. First, we retrieve a set of queries from $\mathcal{D}^{\mathrm{src}}$:
\[
\mathcal{Q}_{j}=\left\{q:\exists (u', q, j, c=1)\in \mathcal{D}^{\mathrm{src}}\right\}.
\]
After that, the queries in $\mathcal{Q}_{j}$ can be ranked according to, for example, the number of clicks on item $j$ in $\mathcal{D}^{\mathrm{src}}$. The top-$N$ queries are kept, denoted by $\{q_{k}\}_{k=1}^N \subseteq \mathcal{Q}_{j}$. Finally, IVs for the item $j$, therefore, can be defined as a stack of the embeddings of the top-$N$ queries:
\begin{equation*}
    \mathbf{Z}_{j} = \left[\mathbf{t}_{q_1},\cdots,\mathbf{t}_{q_k},\cdots,\mathbf{t}_{q_N} \right],
\end{equation*}
where $\mathbf{Z}_j\in \mathbb{R}^{d_q\times N}$, and $\mathbf{t}_{q_k}\in \mathbb{R}^{d_q}$ is the embedding vector of $q_k$.\footnote{To ensure $Z_j$ is a $d_q\times N$ matrix, we recall several similar query embeddings to be inserted to the right side of $\mathbf{Z}_{j}$ if $|\mathcal{Q}_j| < N$. More details can be found in the experiment.} The query embeddings can be obtained by the model of BERT. 

As pre-processing, for each item $j$ in search data $\mathcal{D}^{\mathrm{src}}$, 
relevant queries $\mathcal{Q}_{j}$ are collected and stacked to compose the IV $Z_{j}$ offline.

\subsection{\mbox{Treatment reconstruction}}\label{sec:treatReconstruct}
Based on the original treatment $\mathcal{T}_{u,i}$ and IVs $\mathcal{Z}_{u,i}$, we show that a new treatment $\mathcal{T}_{u,i}^{\mathrm{re}}$ can be created by first regressing $\mathcal{T}_{u,i}$ on $\mathcal{Z}_{u,i}$ and then combining the fitted vectors and the residuals, shown in right part of  \autoref{fig:framework}.

\subsubsection{Treatment decomposition}
The goal of IVs method is to isolate the causal association flowing from the treatments to outputs. As shown in~\autoref{fig:overall}(b), according to the attributes of IVs (i.e. IVs are unconfounded by the confounder and only affect the outcome $Y$ via treatment $X$), we regress $\mathcal{T}_{u,i}$ on $\mathcal{Z}_{u,i}$ to get $\widehat{\mathcal{T}}_{u,i}$ which doesn't depend on the confounders $B$:
\begin{equation}
\begin{split}
      \widehat{\mathcal{T}}_{u,i} = &\left\{\hat{\mathbf{t}}_j= f_{\textrm{proj}}(\mathbf{t}_j, \mathbf{Z}_j):  j\in \mathcal{I}_u \cup \{i\}\right\}, \\
\end{split}
\end{equation}
where $\mathbf{t}_j \in \mathcal{T}_{u,i}$ and $\mathbf{Z}_j\in \mathcal{Z}_{u,i}$ and $f_{\mathrm{proj}}:\mathbb{R}^{d_i}\times \mathbb{R}^{d_q\times N} \mapsto \mathbb{R}^{d_q}$ is defined as a product of matrix $\mathbf{Z}_j$ with an $N$-dimensional vector $\tau_j$:
\[
f_{\mathrm{proj}}(\mathbf{t}_j,\mathbf{Z}_j)=  \mathbf{Z}_j \mathbf{\tau}_j,
\]
where $\tau_j$ is a closed form solution of a least square regression:
\begin{equation*}
    \tau_j = \argmin\limits_{\tau_j\in\mathbf{R}^{N}}
    \left\| \mathbf{Z}_j\tau_j - \textrm{MLP}_0(\mathbf{t}_j) \right\|_2^2
     =\mathbf{Z}_j^\dagger \textrm{MLP}_0(\mathbf{t}_j),
\end{equation*}
where $\mathbf{Z}_j^\dagger$ is the Moore-Penrose pseudoinverse of $\mathbf{Z}_j$ and
$\textrm{MLP}_0: \mathbb{R}^{d_i} \mapsto \mathbb{R}^{d_q}$ is a multi-layer Perceptron that maps the item $j$'s embedding to a latent space of $d_q$ dimensions. 
We call $\hat{\mathbf{t}}_j \in \widehat{\mathcal{T}}_{u,i}$ the \emph{fitted part} of the embedding $\mathbf{t}_j$, which reflects the causal association between the embedding and the outcome in RS. 

After getting the fitted vectors in $\widehat{\mathcal{T}}_{u,i}$, it is easy to get the \emph{residual part} of the regression $\tilde{\mathcal{T}}_{u,i}$:
\begin{equation}\label{eq:residule}
\tilde{\mathcal{T}}_{u,i} = \left\{\tilde{\mathbf{t}}_j = \textrm{MLP}_0(\mathbf{t}_j) - \hat{\mathbf{t}}_j:  j\in \mathcal{I}_u \cup \{i\}\right\}, 
\end{equation}
which contains the non-causal association in the RS. 
The intuition is that the nonlinear representation of the embedding is projected onto the subspace spanned by the columns of the IVs, which separates the fitted part from the residual part. 
Intervening the fitted part and the residual part differently could help mining the different mechanisms of these two parts for outcome prediction in RS. 

Please note that the traditional methods of IVs usually make use of linear models for the regression. Here the linearity assumption is relaxed by first mapping the treatments into a latent space with a nonlinear neural network, which makes our method enjoys both the properties of IVs methods and the powerful representation ability of nonlinear neural networks. 

\subsubsection{Treatment combination}
The fitted vectors $\widehat{\mathcal{T}}_{u,i}$ and the residuals $\tilde{\mathcal{T}}_{u,i}$ can be recombined, achieving a reconstructed treatment:
\begin{equation}\label{eq:reconstructV}
\mathcal{T}^{\mathrm{re}}_{u,i} = \left\{\mathbf{t}^{\mathrm{re}}_j = \alpha^1_j \hat{\mathbf{t}}_j + \alpha^2_j \tilde{\mathbf{t}}_j: j\in \mathcal{I}_u \cup \{i\}\right\},
\end{equation}
where $\hat{\mathbf{t}}_j \in \widehat{\mathcal{T}}_{u,i}$ and $\tilde{\mathbf{t}}_j \in \widetilde{\mathcal{T}}_{u,i}$ are the vectors in these two sets, both correspond to the same item $j$, and $\alpha_j^1\in \mathbb{R}$ and $\alpha_j^2\in \mathbb{R}$ are two combination weights which are estimated by two MLPs:
\[
\alpha_j^1 =  \text{MLP}_1( \textrm{MLP}_0(\mathbf{t}_j), \mathbf{Z}_j); \quad
\alpha_j^2 = \text{MLP}_2( \textrm{MLP}_0(\mathbf{t}_j), \mathbf{Z}_j),
\]
where the inputs of the two different MLPs are concatenations of the transformed $\mathbf{t}_j$ and $\mathbf{Z}_j$ corresponding to item $j$. 

In traditional causal inference, the major challenge is how to identify the causal association from observed data. Therefore, the residual part is often discarded to remove the effects from confounders. That is, removing the edge from the confounders $B$ to the residual $\tilde{\mathcal{T}}_{u,i}$ in~\autoref{fig:overall}(b). 
In recommendation scenario, however, we still focus on promoting the accuracy of preference estimation, rather than just
identifying cause-effects. Existing studies also found that the non-causal associations can contribute to the prediction accuracy~\citep{leverage_bias_RecSys}. The observation motivates
us that not all confounders (biases) need to be discarded. The residual can be leveraged to improve the recommendation performance.

\subsection{Model-agnostic application}
\label{applied_rec}
Many RSs~\citep{wu-etal-2019-neural,DIN,xue2019deep} share a similar structure, which we refer to as the underlying model, shown in the left part of  \autoref{fig:framework}. 
Underlying models represent items as embedding vectors, utilize user's historical behaviors to learn user representation, and predict the preference score of $(u, i)$ based on their learned representations. 
Our proposed IV4Rec is a model-agnostic framework that can be implemented over existing RSs that follow this underlying structure by simply adding a treatment reconstruction module for item embedding, shown in \autoref{fig:framework}. 
The procedure follows the causal graph in \autoref{fig:overall}(b), where queries are utilized as IVs to reconstruct the treatments. 

Formally, representations of $(u, i)$ in existing RSs, are denoted as $\mathbf{t}_{u}$ and $\mathbf{t}_{i}$, where $\mathbf{t}_{u}$ can be calculated by aggregating the user's historically interacted items and other context information (e.g, user profile, search history and etc). 
After reconstructing treatments, we can get reconstructed item embedding  $t^{\mathrm{re}}_{i}$ and reconstructed user embedding $t^{\mathrm{re}}_{u}$, where $t^{\mathrm{re}}_{u}$ is calculated as:
\begin{equation}
    \mathbf{t}_{u}^{\mathrm{re}} = f_{\mathrm{user}}(\mathbf{t}_{1}^{\mathrm{re}},\mathbf{t}_{2}^{\mathrm{re}},\cdots,\mathbf{t}_{n}^{\mathrm{re}},\mathbf{p}_{u}),
\end{equation}
where $\mathbf{t}_{1}^{\mathrm{re}},\mathbf{t}_{2}^{\mathrm{re}},\cdots,\mathbf{t}_{n}^{\mathrm{re}}$ are reconstructed item vectors in the set of interacted items $\mathcal{I}_{u}$, $\mathbf{p}_{u}$ is the representation of other context for user $u$ and $f_{\mathrm{user}}$ can be any module that learns user representation from user's behaviors, e.g. attention mechanism in \citep{DIN,wu-etal-2019-neural}. 
Finally, user's preference is predicted based on learned user/item representation:
\begin{equation}
    \hat{y}_{u,i}=f_{\mathrm{pred}}(t^{\mathrm{re}}_{u},t^{\mathrm{re}}_{i}),
\end{equation}
where $f_{\mathrm{pred}}$ can be any model that predicts the preference score from their representations, e.g. MLP~\citep{DIN} or inner product~\citep{wu-etal-2019-neural}.

Please note that the trained treatment reconstruction module can be applied to the items in an offline manner. That is, after the parameters (i.e., the parameters in MLP$_0$, MLP$_1$, and MLP$_2$) in the treatment reconstruction module are determined, the module can be used to reconstruct all of the item embeddings as a pre-processing step. At the online time, the underlying model directly uses the reconstructed items. Therefore, IV4Rec doesn't have any additional time cost at the online recommendation.

\subsection{Model training}
Parameters in the proposed IV4Rec include parameters in MLP$_0$, MLP$_1$, MLP$_2$, and the parameters from the underlying recommendation model. All these trainable parameters are denoted as $\bm \Theta$ and trained based on $\mathcal{D}^{\mathrm{rec}}$. Specifically, the task of model training amounts to optimizing the following cross entropy loss:
\begin{equation}
    \mathcal{L}_{\bm \Theta} = -\frac{1}{|\mathcal{D}^{\textrm{rec}}|}\sum_{(u, i, c)\in \mathcal{D}^{\textrm{rec}}}\!\!\!\!\!\!\!\! c\cdot\text{log}~\hat{y}_{u,i}+(1-c)\cdot\text{log}(1-\hat{y}_{u,i})
    + \lambda \left\| \bm \Theta \right\|^2,
\end{equation}
where 
$\hat{y}_{u,i}$ is the predicted preference socre for $(u, i)$, $\left\| \bm \Theta \right\|^2$ is the regularizer term for avoiding over-fitting, and $\lambda > 0$ is a coefficient.

\section{Discussion}
\subsection{Feasibility of using search queries as IVs}\label{sec:FeasiofIV}
According to the theory of IVs estimations, the IVs have two assumptions: exogeneity and relevance. 

As for exogeneity, it means that the IVs (search queries) are uncorrelated with the (unobservable) confounders. In recommendation, the common confounders are, for example, variant biases including the position biases, selection biases, etc. Note that currently the search and recommendation are usually deployed as two separate services in one app. The queries are issued when the users are conducting search while the biases occur when the users are accessing the service of RS. Also, these queries may be issued by 
the users other than the one who is accessing the RS. Therefore, these search users cannot be influenced by the ranking positions/exposure of the items in the RS. 

As for relevance, it means that the IVs (search queries) are the causes of the treatment, but do not directly affect the outcomes in RS, i.e., user's click behavior. The search and recommendation share a common goal: providing users with items for satisfying their information needs. In search, the users' information needs are explicitly summarized as queries. In RSs, the information needs are implicitly summarized with the representations of users and items. When the search engine and RS are deployed in one app and serve the same group of users with the same set of items, a large extent of search queries reflect some part of the user information needs in recommendation. The phenomenon indicates that search queries can be seen as a cause of the treatment in recommendation. 
Considering that the IVs (search queries associated with an item) are specific requests made by users in search, it is obvious that they do not directly affect the outcomes in RS.

Therefore, we conclude that the embeddings of the search queries satisfy the assumptions of exogeneity and relevance well. 

\subsection{Difference with traditional IVs methods}
In the field of causal inference, IVs methods provide a very powerful framework for learning 
cause-effects between treatments and outcome even in the presence of confounders. 
The proposed IV4Rec, inspired by the IVs methods, enjoys a number of merits from IVs, including the elegant approach to involving external search information for constructing IVs for recommendation, the least square regression for decomposing the treatment. However, 
IV4Rec has made the following fundamental modifications for adapting the traditional method of IVs to recommendation.
    
(1) \textbf{Representation of treatment}: We take the origin embedding as the input of deep neural network and obtain a neural representation of the embedding, rather than directly applying the least square regression. In particular, we update this neural representation by minimizing the loss of the CTR prediction in the recommendation task, making our application of IVs as an end-to-end process. The modification makes the proposed model enjoys the advantages from both IVs and neural networks.    

(2) \textbf{Reconstruction of treatment with both causal and non-causal parts}: In traditional IVs methods, the residual of the least square regression is discarded. In our approach, however, the residual is used as the embedding representation of the indirect association part. This is because our goal is not just identifying the causal associations. Finding a suitable reconstruction of the causal part and the non-causal part from the original treatment is more helpful in enhancing the recommendation accuracy.

In the recommendation task, biases are ubiquitous, e.g., selection bias and popularity bias, while these biases are usually mixed and difficult to identify.  
In this paper, we do not model the biases explicitly, and focus on improving the recommendation performance using search data as IVs, which explores the causal relationship between the search and recommendation tasks. Since IVs can be used to adjust for both observed and unobserved confounding effects, the proposed model can be considered as a causal learning framework for recommendation using search data.

\section{Experiments}

\begin{table*}[t]
\caption {Performance comparisoins of IV4Rec and the baselines on the Kuaishou dataset and the MIND dataset. $*$ and $\dagger$ respectively indicate the improvements over NRHUB and DIN are statistically significant($p-$value $<$ 0.05)}\label{all results} 
\small{
\begin{tabular}{l|cccc|cccc}
\toprule[0.85pt]
\multirow{2}{*}{Model} & \multicolumn{4}{|l}{\hspace{4em} Kuaishou Dataset} & \multicolumn{4}{|l}{\hspace{4em} MIND Dataset}    \\
                       & AUC     & MRR     & nDCG@5  & nDCG@10  & AUC    & MRR    & nDCG@5  & nDCG@10 \\ \midrule[0.85pt]
NRHUB                  & 0.6455  & 0.1816  & 0.4347  & 0.4692   & 0.6595& 0.3123 &0.3428& 0.4065  \\
JSR-NRHUB              &   0.6488    &     0.1812    &0.4326&       0.4687   & 0.6660  &  0.3164* & 0.3480*&  0.4117*\\
IV4Rec-NRHUB             & $\bf 0.6574^*$ & $\bf 0.1837^* $ & $\bf 0.4411^*$ & $\bf 0.4774^*$   & $\bf 0.6722^*$ & $\bf 0.3271^*$ & $\bf 0.3609^*$ & $\bf 0.4219^*$ \\ \hdashline
DIN                    & 0.6512  & 0.1833  & 0.4416  & 0.4743   & 0.6851&0.3326  & 0.3680  & 0.4304  \\
JSR-DIN                &     0.6524    &0.1838 &0.4417 &0.4755 & 0.6873   &   0.3315    &      0.3686   &     0.4308    \\ 
IV4Rec-DIN               & $\bf 0.6561^\dagger$  & $\bf 0.1844$  & $\bf 0.4432^\dagger$  & $\bf 0.4779^\dagger$   & $\bf 0.6898^\dagger$ & $\bf 0.3336$  & $\bf 0.3700^\dagger$  & $\bf 0.4326^\dagger$  \\ \bottomrule[0.85pt]
\end{tabular}
}
\end{table*}

We present experimental results in this section.\footnote{Codes available at~\url{https://github.com/Ethan00Si/Instrumental-variables-for-recommendation}.}

\subsection{Experimental settings}

\begin{table}[t]
 \caption {Statistics of datasets used in this paper.}\label{dataset} 
{
\begin{tabular}{lllll}
\hline
 Dataset & User & Item & Query & Interaction \\ \hline
 Kuaishou & 12,000  & 3,053,966 &162,624  &4,001,613   \\
 MIND & 736,349  & 130,380 &130,380  &95,447,571   \\ \hline
\end{tabular}
}
\vspace{-0.5cm}
\end{table}

\subsubsection{Datasets}
IV4Rec requires both search logs and recommendation logs. In the experiments, we created two datasets: one is collected from logs of Kuaishou short-video app, and the other is based on the publicly available MIND dataset~\citep{wu-etal-2020-mind}. 
\autoref{dataset} shows some of the statistics on both datasets.

\textbf{Kuaishou Dataset:} 
The Kuaishou dataset is created based on the activities of 12,000 randomly selected users when they elected to use both the search and recommendation services on an app named Kuaishou\footnote{\url{https://www.kuaishou.com/en}}, one of the largest short-video platforms in China, over a period of 7 days in May 2021. The historical behaviors in search and recommendation services of each user were collected. For each user, item  and query in the dataset, the user context embedding (64 dimensions), item embedding (64 dimensions), and query embedding (64 dimensions) were generated using existing pre-trained and ranking models from the platform. The detail algorithms are omitted due to privacy concerns.

We split the dataset into three subsets in chronological order, i.e., the first 5 days for training, the  6th day for validation, and the last day for testing. 
The mini-batch size is set to be 50.

\textbf{MIND Dataset:} 
To the best of our knowledge, there is no publicly available dataset that contains both user's search and recommendation activities. Therefore, we enhance the MIND\footnote{\url{https://msnews.github.io/}}~\citep{wu-etal-2020-mind} data, a benchmark for news recommendation, by generating queries from its metadata. Specifically, motivated by the observation in~\citep{rowley_2000}, one search query for each news article was created by concatenating the texts of its category, subcategory and the entities in the metadata. For a few number of articles where the entities are missing, 
“NLTK”\footnote{\url{https://www.nltk.org/}} was used to extract entities from the titles. 
To generate the query and item embeddings, we follow~\cite{wu-etal-2020-mind} by using BERT~\citep{BERT} to generate the item embeddings (768 dimensions), where the input is the concatenation of the title and abstract. Query embeddings (768 dimensions) are generated by using query strings as input in the same way. 

Since MIND does not contain a test set with labels, the original training(validation) data is used as training(test) set in the experiments. The mini-batch size is set to be 512.

\subsubsection{Baselines and evaluation metrics}
The proposed IV4Rec is model-agnostic, which can be applied to the following baselines and can improve their performances.\\
\textbf{NRHUB}~\citep{wu-etal-2019-neural}: NRHUB utilizes an attentive multi-view learning framework for news recommendation to aggregate heterogeneous behaviors of users such as search queries, clicked items, and browsed items. On the experiments of MIND dataset, it was adapted by removing the modules using clicked items since only search queries were created on the MIND dataset. On the experiments of Kuaishou dataset, 
news encoder is removed since items are short-videos other than articles.\\
\textbf{DIN}~\citep{DIN}: DIN applies an attention mechanism to mine user interests from historical behaviors w.r.t. a certain candidate item. It was adapted to Kuaishou dataset by adding queries and clicked items in the search history as additional history of user behaviors.

We also compare IV4Rec to JSR~\citep{zamani2018joint} that jointly optimizes search and recommendation. JSR is a general joint training framework that trains a separate search model and recommendation model by optimizing a joint loss. The search component of JSR was designed as a fully-connected feed-forward network, following original paper. The recommendation component was set as NRHUB or DIN, leading to two versions of JSR: \textbf{JSR-NRHUB} and \textbf{JSR-DIN}. 

The proposed IV4Rec is model-agnostic. In the experiments, we applied IV4Rec to the following baselines, achieving two versions of our approach, referred to as \textbf{IV4Rec-DIN} and \textbf{IV4Rec-NRHUB}.

As for evaluation metrics, AUC was adopted to measure the prediction accuracy on the clicks. MRR and nDCG at the positions of 5, and 10 were also used to measure the accuracy of item rankings, using the clicks as relevance labels. 
We reported the average results in AUC, MRR, nDCG@5 and nDCG@10 of all impressions. 

\subsubsection{Implementation details} \label{implementation_detail}
The hyper-parameters of neural networks were optimized using grid search. The learning rate was selected from $\{1e-4, 3e-4, 5e-4, 7e-4, 1e-3\}$ and the dropout keep probability was selected from $\{0.5, 0.9, 1.0\}$. 
For the baselines, we set the parameters as the optimal values reported in the original paper. 
Adam~\citep{DBLP:journals/corr/KingmaB14} is used to conduct the optimization. 

As described in section \ref{const_V_Z}, top $N$ associated queries were used to construct the IVs. 
$N$ was set to $10$ on the Kuaishou dataset. 
In~\autoref{dataset}, the query-click data is very sparse and most items have few associated search clicks. To overcome the sparsity problem, we leveraged cosine similarity of the item embedding and the query embedding to measure the strength of the association. Query-item pairs with high cosine similarity were used as complementary to the sparse query-click data. On the MIND dataset, $N$ was set to 1 since only one query was created for each item (news article). 

\subsection{Experimental results}

From the results reported in \autoref{all results}, we found that IV4Rec-NRHUB and IV4Rec-DIN significantly outperformed the corresponding underlying models, NRHUB and DIN, on both datasets, with statistical significance. The results verified the effectiveness of the model-agnostic IV4Rec framework in improving any recommendation models. On the other hand, IV4Rec-NRHUB and IV4Rec-DIN also outperformed the baselines of JSR-NRHUB and JSR-DIN, which jointly optimize search and recommendation. 
Please note that NRHUB leveraged search activities for user modeling and DIN was adapted by adding users' search histories on the Kuaishou dataset. 
Thus the improvements achieved on the Kuaishou dataset were attributed to IV4Rec instead of adding users' search history features.
The results verified the effects of using search queries as IV for reconstructing treatments in recommendation.

\subsection{Detailed empirical analysis}
We conducted more detailed experiments to show how and why IV4Rec can improve the recommendation accuracy. 

\subsubsection{Effects of search queries as IVs.} 
To verify the relevance assumption in Section~\ref{sec:FeasiofIV}, 
experiments were conducted on the Kuaishou dataset. 
\begin{figure}[t]
        \centering
            \includegraphics[width=0.30\textwidth]{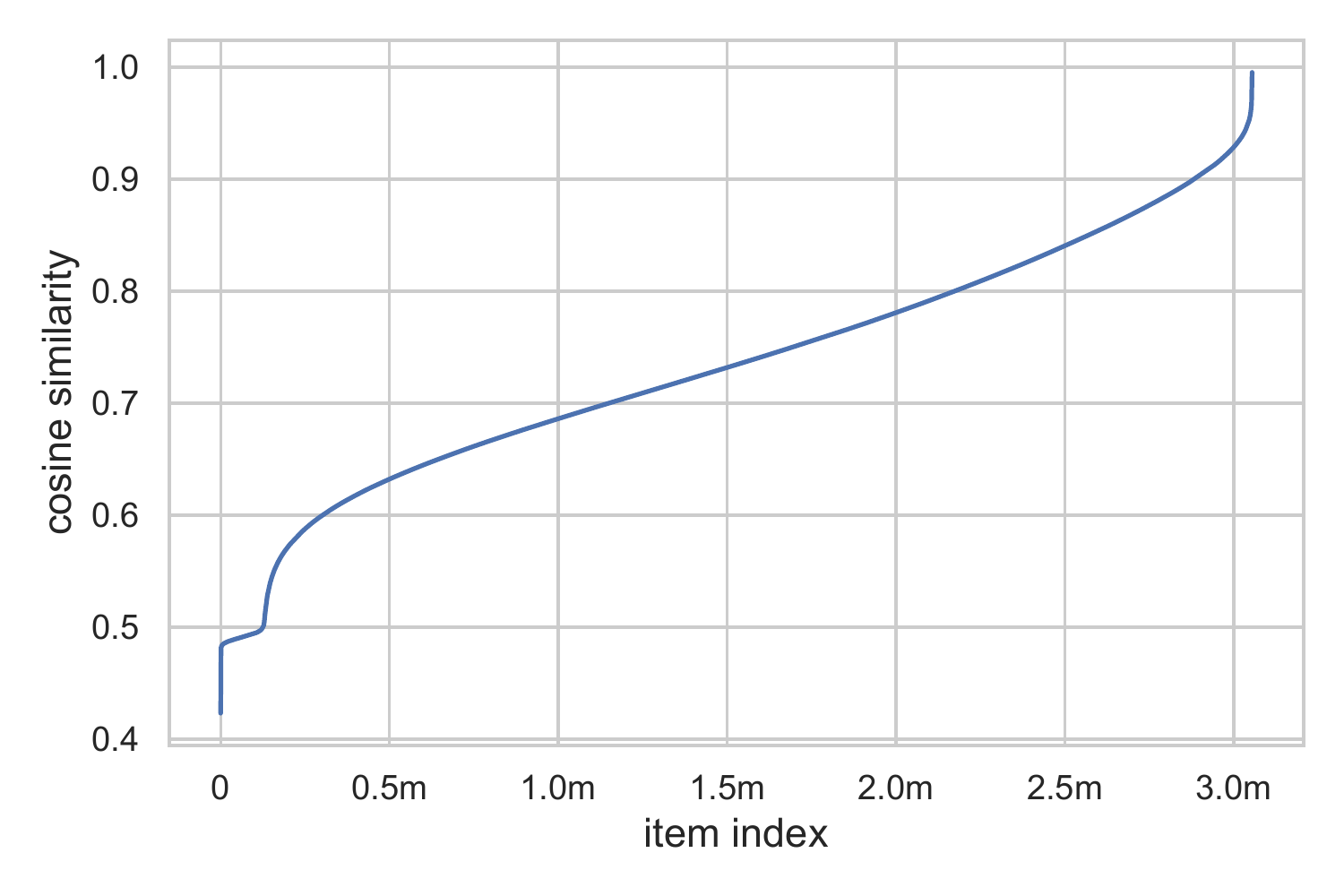}
         \caption{Distribution of cosine similarity between each item and its highest ranked query. The items are sorted by cosine similarity from the lowest to the highest. Each x-axis index refers to a unique item.}
         \label{fig:relevance of IVs}
\vspace{-0.5cm}
\end{figure}
Specifically, we tested the relevance between items and their corresponding queries. As discussed in \ref{implementation_detail}, we used cosine similarity of query-item pair embeddings to measure the relevance. The embeddings of queries and items were generated using pretrained models from the platform. The similarity of each item and its highest ranked query was plotted in \autoref{fig:relevance of IVs}. From the results, most similarity scores were higher than 0.6, indicating that most items were highly relevant with corresponding queries.

\begin{figure}
\subfigure[AUC w.r.t. \# clicked queries]{\includegraphics[width=0.23\textwidth]{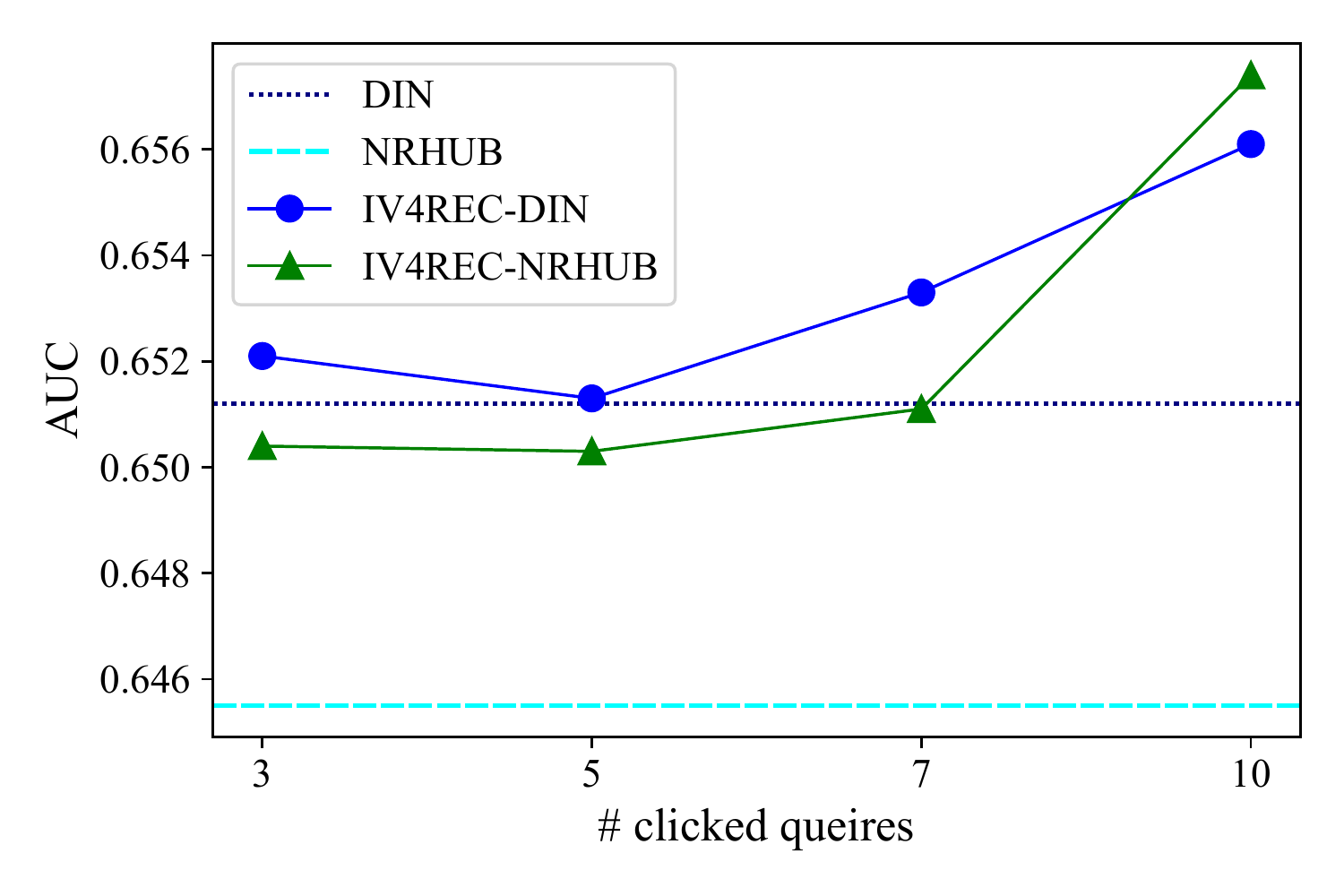}}
\subfigure[AUC w.r.t. fraction of clicked queries]{\includegraphics[width=0.23\textwidth]{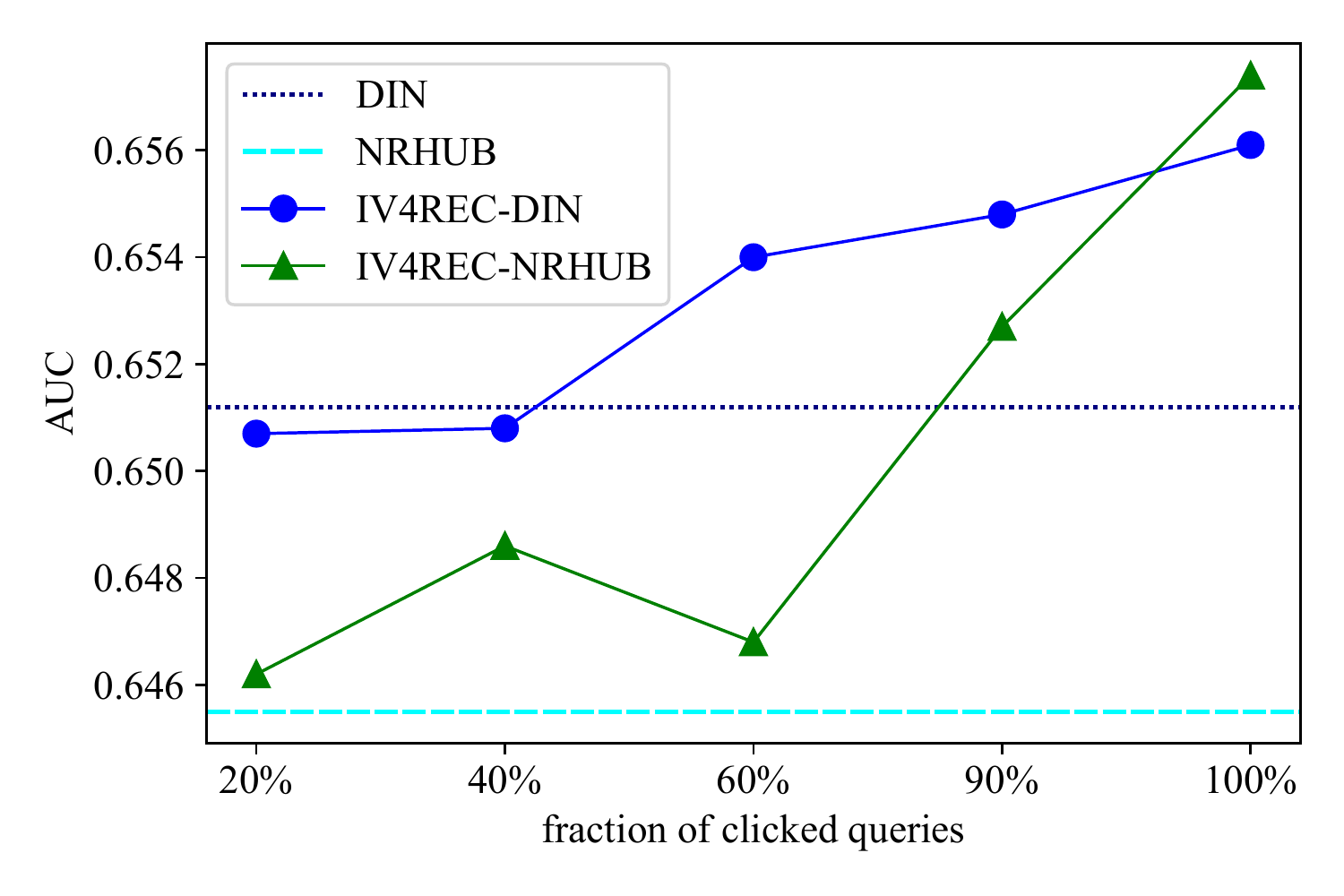}}
\caption{AUC curves when the IVs (search queries) are selected differently on Kuaishou dataset. Two horizontal lines denotes the performances of DIN and NRHUB, respectively. IV4REC significantly outperforms the baselines when each item has more than 5 relevant queries.}
\vspace{-2ex}
\label{fig:exp on IVs}
\end{figure}

We explored the impacts of the number of relevant queries per item. Specifically, we tested the performances of IV4Rec when the number of queries extracted for each item (i.e., $N$ in $\mathbf{Z}_j\in \mathbb{R}^{d_q\times N}$) as the IVs.~\autoref{fig:exp on IVs}(a) showed the AUC curves of IV4Rec models w.r.t. $N = 3, 5, 7, 10$. We found that with the increased number of $N$ (more related queries means higher relevance between IV and treatment), AUC also increased for both IV4Rec-NRHUB and IV4Rec-DIN. 

We also tested the performances of IV4Rec when a few queries in the IVs were selected randomly rather than the high-ranked queries according to the clicks in search. \autoref{fig:exp on IVs}(b) illustrates the AUC curves w.r.t. 20\%, 40\%, 60\%, 80\% and 100\% of the queries are clicked queries (others are random queries). From the results, we can see that when more queries in IVs were set randomly (lower relevance of IVs to treatments), more hurts to the performances of IV4Rec-NRHUB and IV4Rec-DIN. 
Based on the results, we conclude that the clicked search queries are effective IVs for recommendation. 

\subsubsection{Effects of using residuals in recommendation.} 
\begin{figure}[t]
        \centering
            \includegraphics[width=0.45\textwidth]{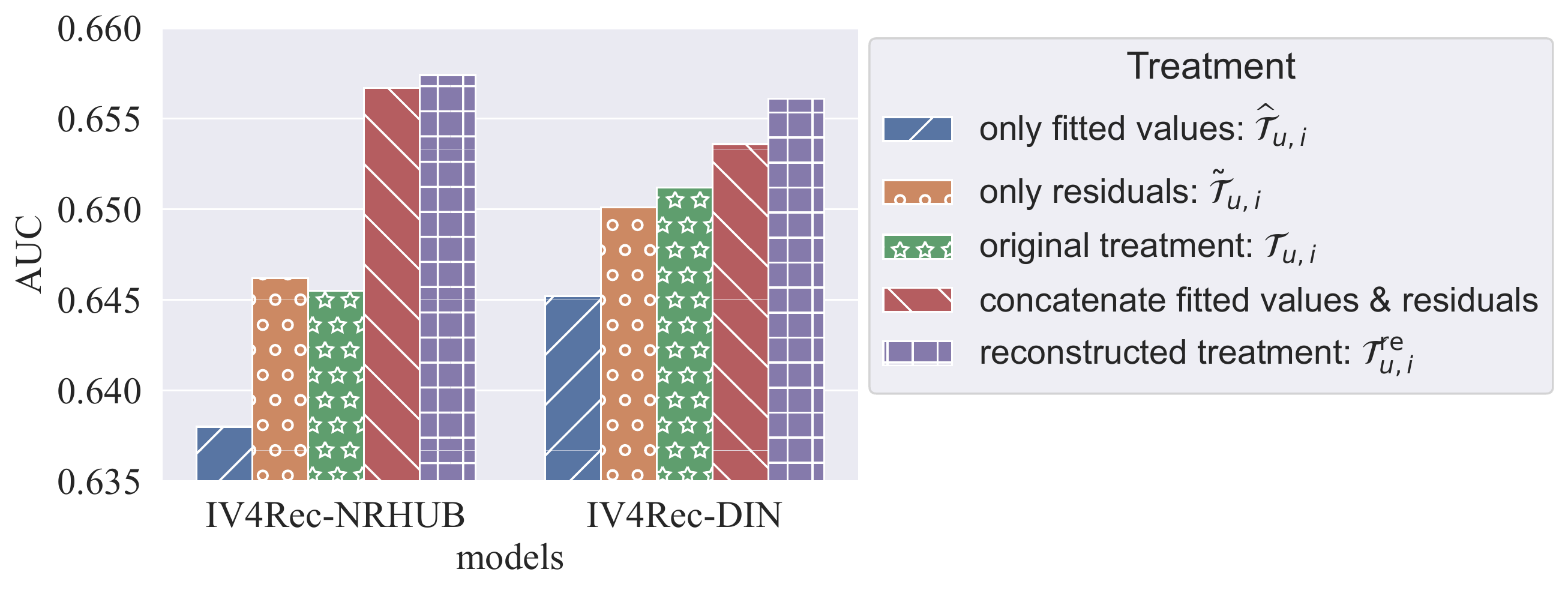}
         \caption{Impact of different treatment reconstruction methods w.r.t. AUC on the Kuaishou dataset.}
         \label{fig:reconstruct}
\vspace{-0.5cm}
\end{figure}
In traditional IV estimation (e.g., 2SLS~\citet{kmenta2010mostly}), the residuals are discarded. In IV4Rec, we utilized both the fitted part and the residual part. 
Experiments were done to test the AUC of different modified IV4Rec versions on the Kuaishou dataset. They are using only the fitted values $\hat{\mathcal{T}}_{u,i}$, using only the residuals $\tilde{\mathcal{T}}_{u,i}$, using the original treatment $\mathcal{T}_{u,i}$ without reconstruction, using the reconstructed treatment by concatenating $\hat{\mathcal{T}}_{u,i}$ and $\tilde{\mathcal{T}}_{u,i}$, using $\mathcal{T}^{re}_{u,i}$ reconstructed by IV4Rec. From the results shown in~\autoref{fig:reconstruct}, we can see that both the fitted part and the residual part have contributions in recommendation. The AUC improved a lot when these two parts are combined together as a reconstructed treatment. The phenomenon can be observed when both NRHUB and DIN were used as the underlying model of IV4Rec. The results indicate that though they represent the non-causal associations, the residual part can still contribute to the user preference prediction. The reason is that the residuals still have a strong association with the outcome. When the goal is making accurate prediction rather than analyzing the causal effects, the fitted part and the residual part are complementary.

Compared to the two versions of combination, the proposed IV4Rec, which uses weighted combination and two MLPs to estimate the weights, performed better than the simple concatenation. The results verified the effectiveness of the treatment reconstruction method in Section~\ref{sec:treatReconstruct}.

\subsubsection{Enhancing the item embeddings.} 
\begin{figure}
\subfigure[Original embeddings by BERT ]{\includegraphics[width=0.23\textwidth]{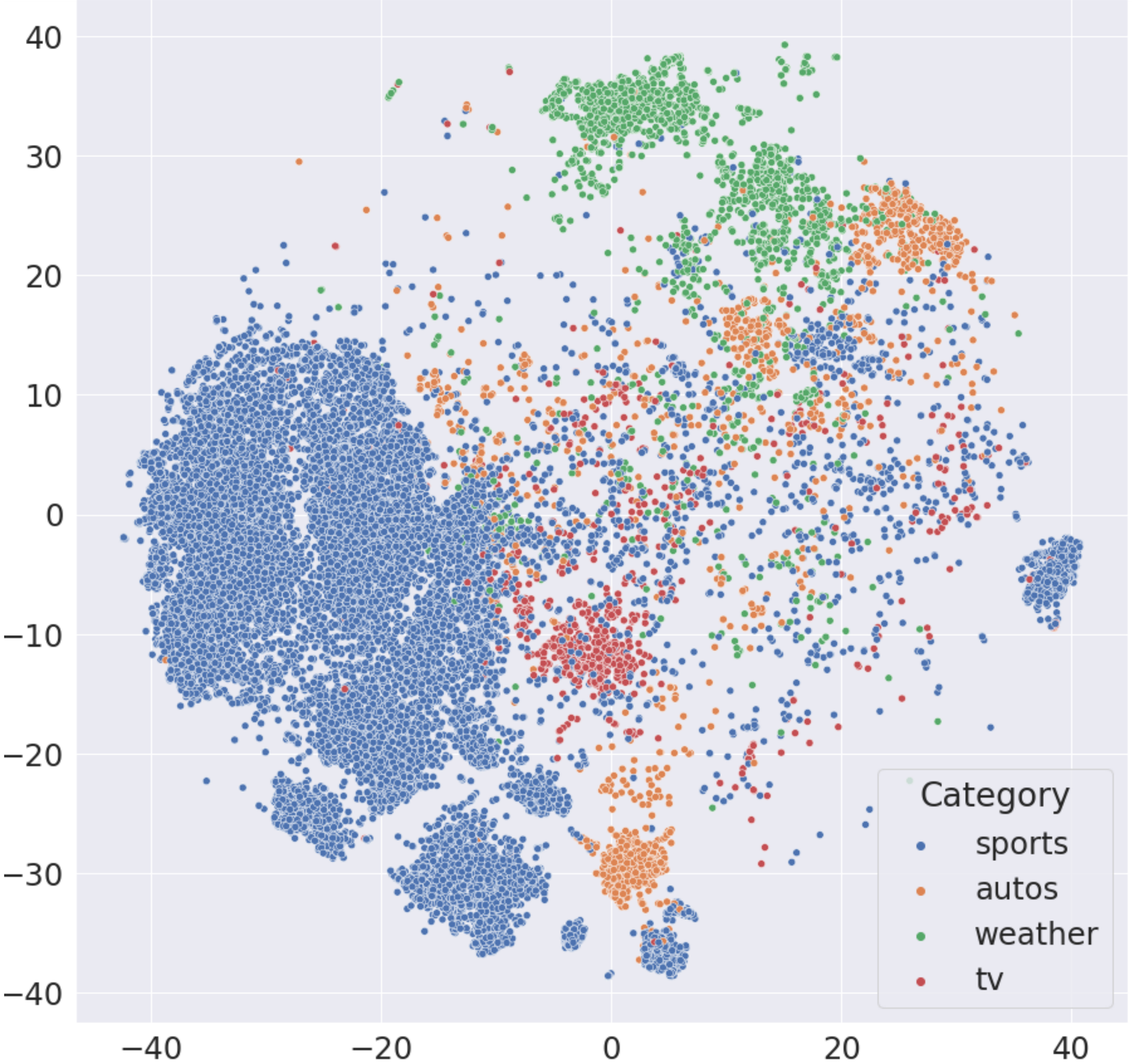}}
\subfigure[Reconstructed embeddings in IV4Rec-NRHUB]{\includegraphics[width=0.23\textwidth]{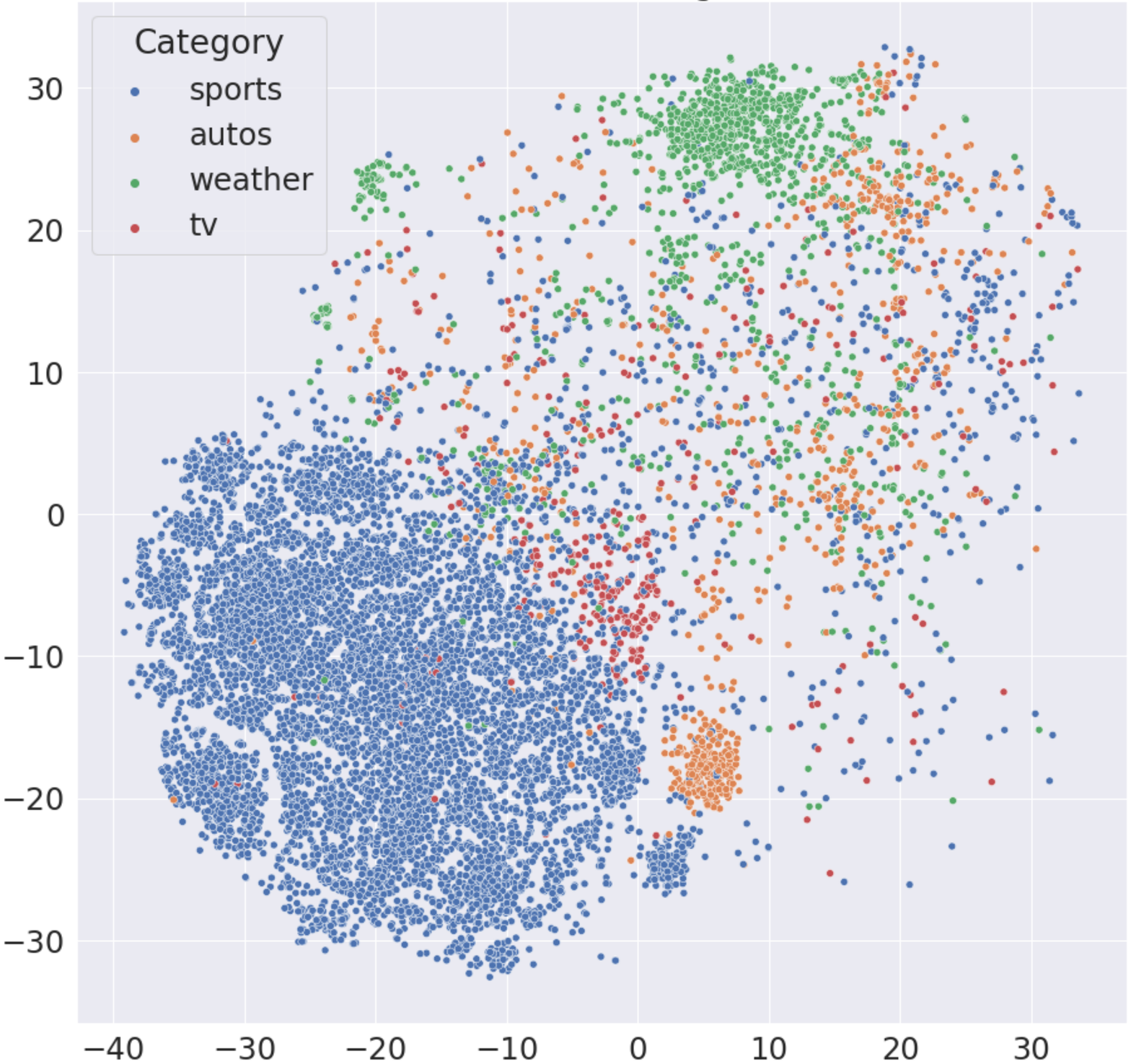}}
\caption{Visualization of the item embeddings on MIND dataset. Using IV4Rec can better cluster embeddings within the same category.}
\label{embedding}
\vspace{-0.5cm}
\end{figure}
We conducted experiments to illustrate whether the reconstructed treatments are better item embeddings than the original ones. The experiments were conducted based on MIND dataset because each news article in MIND falls into a category. We selected the articles from four categories (sports, autos, weather, and tv), and illustrated their original embeddings $\mathbf{t}_j$'s (by BERT) in Figure~\ref{embedding}(a) with t-SNE~\citep{van2008visualizing} where the four colors indicate four categories. Based on IV4Rec-NRHUB, we also calculated the reconstructed item embeddings $\mathbf{t}_j^{re}$'s of these news articles, and illustrated them in Figure~\ref{embedding}(b). Comparing these two figures, we found that the reconstructed embeddings are distributed better than the original embeddings. For example, the `sports' articles are more tightly clustered at the bottom-left corner. The results indicate that IV4Rec has the ability to improve the item embeddings with the help of search queries. It also provides an explanation of why IV4Rec can improve the underlying model.

\section{Conclusions}
In this paper, we proposed a model agnostic IV-based causal learning framework to improve recommendation using search data, called IV4Rec.
IV4Rec made use of the search queries as IVs and  decomposed the recommendation embeddings into the causal association part and the non-causal association part, mining the different mechanisms of these two parts for preference prediction in recommendation. 
Besides, IV4Rec combined the traditional method of instrumental variables with deep neural networks and provided an end-to-end framework for estimating the model parameters. 
Experiments on Kuaishou product data and a public benchmark demonstrated the effectiveness of IV4Rec in recommendation. 

%
\begin{acks}
This work was funded by the National Key R\&D Program of China (2019YFE0198200), Kuaishou,
the National Natural Science Foundation of China (61872338, 62006234, 61832017), Project funded by China Postdoctoral Science Foundation (No. 2021T140722), Beijing Outstanding Young Scientist Program 
NO. BJJWZYJH012019100020098, Intelligent Social Governance Interdisciplinary Platform, 
Major Innovation \& Planning Interdisciplinary Platform for the ``Double-First Class'' Initiative, 
Renmin University of China, and Public Policy and Decision-making Research Lab of Renmin University of China.
\end{acks}
%
%



\balance
\bibliographystyle{ACM-Reference-Format}
\bibliography{www}

\end{document}